%
%
%
\def\unredoffs{} \def\redoffs{\voffset=-.31truein\hoffset=-.48truein}
\def\speclscape{}
%
%
%
%
%
\newbox\leftpage \newdimen\fullhsize \newdimen\hstitle \newdimen\hsbody
\tolerance=1000\hfuzz=2pt
\catcode`\@=11 
\ifx\hyperdef\UNd@FiNeD\def\hyperdef#1#2#3#4{#4}\def\hyperref#1#2#3#4{#4}\fi
\def\bigans{b }
\def\answ{b }
%
\ifx\answ\bigans\message{(This will come out unreduced.}
\magnification=1200\unredoffs\baselineskip=16pt plus 2pt minus 1pt
\hsbody=\hsize \hstitle=\hsize 
\else\message{(This will be reduced.} \let\l@r=L
\magnification=1000\baselineskip=16pt plus 2pt minus 1pt \vsize=7truein
\redoffs \hstitle=8truein\hsbody=4.75truein\fullhsize=10truein\hsize=\hsbody
\output={\ifnum\pageno=0 
  \shipout\vbox{\speclscape{\hsize\fullhsize\makeheadline}
    \hbox to \fullhsize{\hfill\pagebody\hfill}}\advancepageno
  \else
  \almostshipout{\leftline{\vbox{\pagebody\makefootline}}}\advancepageno
  \fi}
\def\almostshipout#1{\if L\l@r \count1=1 \message{[\the\count0.\the\count1]}
      \global\setbox\leftpage=#1 \global\let\l@r=R
 \else \count1=2
  \shipout\vbox{\speclscape{\hsize\fullhsize\makeheadline}
      \hbox to\fullhsize{\box\leftpage\hfil#1}}  \global\let\l@r=L\fi}
\fi
%
\newcount\yearltd\yearltd=\year\advance\yearltd by -2000

\def\Title#1#2{\nopagenumbers\abstractfont\hsize=\hstitle\rightline{#1}%
\vskip 1in\centerline{\titlefont #2}\abstractfont\vskip .5in\pageno=0}
\def\Date#1{\vfill\leftline{#1}\tenpoint\supereject\global\hsize=\hsbody%
\footline={\hss\tenrm\hyperdef\hypernoname{page}\folio\folio\hss}}%
%

\def\draftmode{\message{ DRAFTMODE }\def\draftdate{{\rm preliminary draft:
\number\month/\number\day/\number\yearltd\ \ \hourmin}}%
\headline={\hfil\draftdate}\writelabels\baselineskip=20pt plus 2pt minus 2pt
 {\count255=\time\divide\count255 by 60 \xdef\hourmin{\number\count255}
  \multiply\count255 by-60\advance\count255 by\time
  \xdef\hourmin{\hourmin:\ifnum\count255<10 0\fi\the\count255}}}
\def\nolabels{\def\wrlabeL##1{}\def\eqlabeL##1{}\def\reflabeL##1{}}
\def\writelabels{\def\wrlabeL##1{\leavevmode\vadjust{\rlap{\smash%
{\line{{\escapechar=` \hfill\rlap{\sevenrm\hskip.03in\string##1}}}}}}}%
\def\eqlabeL##1{{\escapechar-1\rlap{\sevenrm\hskip.05in\string##1}}}%
\def\reflabeL##1{\noexpand\llap{\noexpand\sevenrm\string\string\string##1}}}
\nolabels
%
\global\newcount\secno \global\secno=0
\global\newcount\meqno \global\meqno=1
\def\s@csym{}
\def\newsec#1{\global\advance\secno by1%
{\toks0{#1}\message{(\the\secno. \the\toks0)}}%
\global\subsecno=0\eqnres@t\let\s@csym\secsym\xdef\secn@m{\the\secno}\noindent
{\bf\hyperdef\hypernoname{section}{\the\secno}{\the\secno.} #1}%
\writetoca{{\string\hyperref{}{section}{\the\secno}{\it\the\secno.}} {{\it #1} }}%
\par\nobreak\medskip\nobreak}
\def\eqnres@t{\xdef\secsym{\the\secno.}\global\meqno=1\bigbreak\bigskip}
\def\sequentialequations{\def\eqnres@t{\bigbreak}}\xdef\secsym{}
\global\newcount\subsecno \global\subsecno=0
\def\subsec#1{\global\advance\subsecno by1%
{\toks0{#1}\message{(\s@csym\the\subsecno. \the\toks0)}}%
\ifnum\lastpenalty>9000\else\bigbreak\fi       \global\subsubsecno=0
\noindent{\it\hyperdef\hypernoname{subsection}{\secn@m.\the\subsecno}%
{\secn@m.\the\subsecno.} #1}\writetoca{\string\quad
{\string\hyperref{}{subsection}{\secn@m.\the\subsecno}{\secn@m.\the\subsecno.}}
{#1}}\par\nobreak\medskip\nobreak}
\def\appendix#1#2{\global\meqno=1\global\subsecno=0\xdef\secsym{\hbox{#1.}}%
\bigbreak\bigskip\noindent{\bf Appendix \hyperdef\hypernoname{appendix}{#1}%
{#1.} #2}{\toks0{(#1. #2)}\message{\the\toks0}}%
\xdef\s@csym{#1.}\xdef\secn@m{#1}%
\writetoca{\string\hyperref{}{appendix}{#1}{{\it Appendix} {\it #1.}} {\it #2}}%
\par\nobreak\medskip\nobreak}
%
%
\def\checkm@de#1#2{\ifmmode{\def\f@rst##1{##1}\hyperdef\hypernoname{equation}%
{#1}{#2}}\else\hyperref{}{equation}{#1}{#2}\fi}
\def\eqnn#1{\DefWarn#1\xdef #1{(\noexpand\relax\noexpand\checkm@de%
{\s@csym\the\meqno}{\secsym\the\meqno})}%
\wrlabeL#1\writedef{#1\leftbracket#1}\global\advance\meqno by1}
\def\f@rst#1{\c@t#1a\em@ark}\def\c@t#1#2\em@ark{#1}
\def\eqna#1{\DefWarn#1\wrlabeL{#1$\{\}$}%
\xdef #1##1{(\noexpand\relax\noexpand\checkm@de%
{\s@csym\the\meqno\noexpand\f@rst{##1}}{\hbox{$\secsym\the\meqno##1$}})}
\writedef{#1\numbersign1\leftbracket#1{\numbersign1}}\global\advance\meqno by1}
\def\eqn#1#2{\DefWarn#1%
\xdef #1{(\noexpand\hyperref{}{equation}{\s@csym\the\meqno}%
{\secsym\the\meqno})}$$#2\eqno(\hyperdef\hypernoname{equation}%
{\s@csym\the\meqno}{\secsym\the\meqno})\eqlabeL#1$$%
\writedef{#1\leftbracket#1}\global\advance\meqno by1}
\def\xeqn{\expandafter\xe@n}\def\xe@n(#1){#1}
\def\xeqna#1{\expandafter\xe@n#1}
\def\eqns#1{(\e@ns #1{\hbox{}})}
\def\e@ns#1{\ifx\UNd@FiNeD#1\message{eqnlabel \string#1 is undefined.}%
\xdef#1{(?.?)}\fi{\let\hyperref=\relax\xdef\next{#1}}%
\ifx\next\em@rk\def\next{}\else%
\ifx\next#1\xeqn#1\else\def\n@xt{#1}\ifx\n@xt\next#1\else\xeqna#1\fi
\fi\let\next=\e@ns\fi\next}

\def\DefWarn#1{\ifx\UNd@FiNeD#1\else
\immediate\write16{*** WARNING: the label \string#1 is already defined ***}\fi}
%
\newskip\footskip\footskip14pt plus 1pt minus 1pt 
\def\footnotefont{\ninepoint}\def\f@t#1{\footnotefont #1\@foot}
\def\f@@t{\baselineskip\footskip\bgroup\footnotefont\aftergroup\@foot\let\next}
\setbox\strutbox=\hbox{\vrule height9.5pt depth4.5pt width0pt}
\global\newcount\ftno \global\ftno=0
\def\foot{\global\advance\ftno by1\def\foot@rg{\hyperref{}{footnote}%
{\the\ftno}{\the\ftno}\xdef\foot@rg{\noexpand\hyperdef\noexpand\hypernoname%
{footnote}{\the\ftno}{\the\ftno}}}\footnote{$^{\foot@rg}$}}
%
\newwrite\ftfile
\def\footend{\def\foot{\global\advance\ftno by1\chardef\wfile=\ftfile
\hyperref{}{footnote}{\the\ftno}{$^{\the\ftno}$}%
\ifnum\ftno=1\immediate\openout\ftfile=\jobname.fts\fi%
\immediate\write\ftfile{\noexpand\smallskip%
\noexpand\item{\noexpand\hyperdef\noexpand\hypernoname{footnote}
{\the\ftno}{f\the\ftno}:\ }\pctsign}\findarg}%
\def\footatend{\vfill\eject\immediate\closeout\ftfile{\parindent=20pt
\centerline{\bf Footnotes}\nobreak\bigskip\input \jobname.fts }}}
\def\footatend{}
%
%
\global\newcount\refno \global\refno=1
\newwrite\rfile
\def\ref{[\hyperref{}{reference}{\the\refno}{\the\refno}]\nref}
\def\nref#1{\DefWarn#1%
\xdef#1{[\noexpand\hyperref{}{reference}{\the\refno}{\the\refno}]}%
\writedef{#1\leftbracket#1}%
\ifnum\refno=1\immediate\openout\rfile=\jobname.refs\fi
\chardef\wfile=\rfile\immediate\write\rfile{\noexpand\item{[\noexpand\hyperdef%
\noexpand\hypernoname{reference}{\the\refno}{\the\refno}]\ }%
\reflabeL{#1\hskip.31in}\pctsign}\global\advance\refno by1\findarg}
\def\findarg#1#{\begingroup\obeylines\newlinechar=`\^^M\pass@rg}
{\obeylines\gdef\pass@rg#1{\writ@line\relax #1^^M\hbox{}^^M}%
\gdef\writ@line#1^^M{\expandafter\toks0\expandafter{\striprel@x #1}%
\edef\next{\the\toks0}\ifx\next\em@rk\let\next=\endgroup\else\ifx\next\empty%
\else\immediate\write\wfile{\the\toks0}\fi\let\next=\writ@line\fi\next\relax}}
\def\striprel@x#1{} \def\em@rk{\hbox{}}
\def\lref{\begingroup\obeylines\lr@f}
\def\lr@f#1#2{\DefWarn#1\gdef#1{\let#1=\UNd@FiNeD\ref#1{#2}}\endgroup\unskip}

\def\addref#1{\immediate\write\rfile{\noexpand\item{}#1}} 
\def\listrefs{\footatend\vfill\supereject\immediate\closeout\rfile\writestoppt
\baselineskip=\footskip\centerline{{\bf References}}\bigskip{\parindent=20pt%
\frenchspacing\escapechar=` \input \jobname.refs\vfill\eject}\nonfrenchspacing}
\def\startrefs#1{\immediate\openout\rfile=\jobname.refs\refno=#1}
\def\xref{\expandafter\xr@f}\def\xr@f[#1]{#1}
\def\refs#1{\count255=1[\r@fs #1{\hbox{}}]}
\def\r@fs#1{\ifx\UNd@FiNeD#1\message{reflabel \string#1 is undefined.}%
\nref#1{need to supply reference \string#1.}\fi%
\vphantom{\hphantom{#1}}{\let\hyperref=\relax\xdef\next{#1}}%
\ifx\next\em@rk\def\next{}%
\else\ifx\next#1\ifodd\count255\relax\xref#1\count255=0\fi%
\else#1\count255=1\fi\let\next=\r@fs\fi\next}
%

%
\newwrite\ffile\global\newcount\figno \global\figno=1
\def\fig{fig.~\hyperref{}{figure}{\the\figno}{\the\figno}\nfig}
\def\nfig#1{\DefWarn#1%
\xdef#1{fig.~\noexpand\hyperref{}{figure}{\the\figno}{\the\figno}}%
\writedef{#1\leftbracket fig.\noexpand~\xfig#1}%
\ifnum\figno=1\immediate\openout\ffile=\jobname.figs\fi\chardef\wfile=\ffile%
{\let\hyperref=\relax
\immediate\write\ffile{\noexpand\medskip\noexpand\item{Fig.\ %
\noexpand\hyperdef\noexpand\hypernoname{figure}{\the\figno}{\the\figno}. }
\reflabeL{#1\hskip.55in}\pctsign}}\global\advance\figno by1\findarg}
\def\listfigs{\vfill\eject\immediate\closeout\ffile{\parindent40pt
\baselineskip14pt\centerline{{\bf Figure Captions}}\nobreak\medskip
\escapechar=` \input \jobname.figs\vfill\eject}}
\def\xfig{\expandafter\xf@g}\def\xf@g fig.\penalty\@M\ {}
\def\figs#1{figs.~\f@gs #1{\hbox{}}}
\def\f@gs#1{{\let\hyperref=\relax\xdef\next{#1}}\ifx\next\em@rk\def\next{}\else
\ifx\next#1\xfig #1\else#1\fi\let\next=\f@gs\fi\next}
\def\figin{\epsfcheck\figin}\def\figins{\epsfcheck\figins}
\def\epsfcheck{\ifx\epsfbox\UNd@FiNeD
\message{(NO epsf.tex, FIGURES WILL BE IGNORED)}
\gdef\figin##1{\vskip2in}\gdef\figins##1{\hskip.5in}
\else\message{(FIGURES WILL BE INCLUDED)}%
\gdef\figin##1{##1}\gdef\figins##1{##1}\fi}
\def\DefWarn#1{}
\def\figinsert{\goodbreak\midinsert}
\def\ifig#1#2#3{\DefWarn#1\xdef#1{Fig.~\noexpand\hyperref{}{figure}%
{\the\figno}{\the\figno}}\writedef{#1\leftbracket fig.\noexpand~\xfig#1}%
\figinsert\figin{\centerline{#3}}\medskip\centerline{\vbox{\baselineskip12pt
\advance\hsize by -1truein\noindent\wrlabeL{#1=#1}\footnotefont%
{\bf Fig.~\hyperdef\hypernoname{figure}{\the\figno}{\the\figno}:} #2}}
\bigskip\endinsert\global\advance\figno by1}
\newwrite\lfile
{\escapechar-1\xdef\pctsign{\string\%}\xdef\leftbracket{\string\{}
\xdef\rightbracket{\string\}}\xdef\numbersign{\string\#}}
\def\writedefs{\immediate\openout\lfile=\jobname.defs \def\writedef##1{%
{\let\hyperref=\relax\let\hyperdef=\relax\let\hypernoname=\relax
 \immediate\write\lfile{\string\def\string##1\rightbracket}}}}%
\def\writestop{\def\writestoppt{\immediate\write\lfile{\string\pageno
 \the\pageno\string\startrefs\leftbracket\the\refno\rightbracket
 \string\def\string\secsym\leftbracket\secsym\rightbracket
 \string\secno\the\secno\string\meqno\the\meqno}\immediate\closeout\lfile}}
\def\writestoppt{}\def\writedef#1{}
\def\seclab#1{\DefWarn#1%
\xdef #1{\noexpand\hyperref{}{section}{\the\secno}{\the\secno}}%
\writedef{#1\leftbracket#1}\wrlabeL{#1=#1}}
\def\subseclab#1{\DefWarn#1%
\xdef #1{\noexpand\hyperref{}{subsection}{\secn@m.\the\subsecno}%
{\secn@m.\the\subsecno}}\writedef{#1\leftbracket#1}\wrlabeL{#1=#1}}
\def\applab#1{\DefWarn#1%
\xdef #1{\noexpand\hyperref{}{appendix}{\secn@m}{\secn@m}}%
\writedef{#1\leftbracket#1}\wrlabeL{#1=#1}}
\newwrite\tfile \def\writetoca#1{}
\def\leaderfill{\leaders\hbox to 1em{\hss.\hss}\hfill}
\def\writetoc{\immediate\openout\tfile=\jobname.toc
   \def\writetoca##1{{\edef\next{\write\tfile{\noindent ##1
   \string\leaderfill {\string\hyperref{}{page}{\noexpand\number\pageno}%
                       {\noexpand\number\pageno}} \par}}\next}}}
\newread\ch@ckfile
\def\listtoc{\immediate\closeout\tfile\immediate\openin\ch@ckfile=\jobname.toc
\ifeof\ch@ckfile\message{no file \jobname.toc, no table of contents this pass}%
\else\closein\ch@ckfile\centerline{\bf Contents}\nobreak\medskip%
{\baselineskip=18.5pt  \footnotefont
\parskip=2pt\catcode`\@=12\input\jobname.toc
\catcode`\@=12\bigbreak\bigskip}\fi}
\catcode`\@=12 
%
\edef\tfontsize{\ifx\answ\bigans scaled\magstep3\else scaled\magstep4\fi}
\font\titlerm=cmr10 \tfontsize \font\titlerms=cmr7 \tfontsize
\font\titlermss=cmr5 \tfontsize \font\titlei=cmmi10 \tfontsize
\font\titleis=cmmi7 \tfontsize \font\titleiss=cmmi5 \tfontsize
\font\titlesy=cmsy10 \tfontsize \font\titlesys=cmsy7 \tfontsize
\font\titlesyss=cmsy5 \tfontsize \font\titleit=cmti10 \tfontsize
\skewchar\titlei='177 \skewchar\titleis='177 \skewchar\titleiss='177
\skewchar\titlesy='60 \skewchar\titlesys='60 \skewchar\titlesyss='60
\def\titlefont{\def\rm{\fam0\titlerm}
\textfont0=\titlerm \scriptfont0=\titlerms \scriptscriptfont0=\titlermss
\textfont1=\titlei \scriptfont1=\titleis \scriptscriptfont1=\titleiss
\textfont2=\titlesy \scriptfont2=\titlesys \scriptscriptfont2=\titlesyss
\textfont\itfam=\titleit \def\it{\fam\itfam\titleit}\rm}
 \ifx\answ\bigans\else scaled\magstep1\fi
\ifx\answ\bigans\def\abstractfont{\tenpoint}\else
\font\absit=cmti10 scaled \magstep1
\font\abssl=cmsl10 scaled \magstep1
\font\absrm=cmr10 scaled\magstep1 \font\absrms=cmr7 scaled\magstep1
\font\absrmss=cmr5 scaled\magstep1 \font\absi=cmmi10 scaled\magstep1
\font\absis=cmmi7 scaled\magstep1 \font\absiss=cmmi5 scaled\magstep1
\font\abssy=cmsy10 scaled\magstep1 \font\abssys=cmsy7 scaled\magstep1
\font\abssyss=cmsy5 scaled\magstep1 \font\absbf=cmbx10 scaled\magstep1
\skewchar\absi='177 \skewchar\absis='177 \skewchar\absiss='177
\skewchar\abssy='60 \skewchar\abssys='60 \skewchar\abssyss='60
\def\abstractfont{\def\rm{\fam0\absrm}
\textfont0=\absrm \scriptfont0=\absrms \scriptscriptfont0=\absrmss
\textfont1=\absi \scriptfont1=\absis \scriptscriptfont1=\absiss
\textfont2=\abssy \scriptfont2=\abssys \scriptscriptfont2=\abssyss
\textfont\itfam=\absit \def\it{\fam\itfam\absit}\def\footnotefont{\tenpoint}%
\textfont\slfam=\abssl \def\sl{\fam\slfam\abssl}%
\textfont\bffam=\absbf \def\bf{\fam\bffam\absbf}\rm}\fi
\def\tenpoint{\def\rm{\fam0\tenrm}
\textfont0=\tenrm \scriptfont0=\sevenrm \scriptscriptfont0=\fiverm
\textfont1=\teni  \scriptfont1=\seveni  \scriptscriptfont1=\fivei
\textfont2=\tensy \scriptfont2=\sevensy \scriptscriptfont2=\fivesy
\textfont\itfam=\tenit \def\it{\fam\itfam\tenit}\def\footnotefont{\ninepoint}%
\textfont\bffam=\tenbf \def\bf{\fam\bffam\tenbf}\def\sl{\fam\slfam\tensl}\rm}
\font\ninerm=cmr9 \font\sixrm=cmr6 \font\ninei=cmmi9 \font\sixi=cmmi6
\font\ninesy=cmsy9 \font\sixsy=cmsy6 \font\ninebf=cmbx9
\font\nineit=cmti9 \font\ninesl=cmsl9 \skewchar\ninei='177
\skewchar\sixi='177 \skewchar\ninesy='60 \skewchar\sixsy='60
\def\ninepoint{\def\rm{\fam0\ninerm}
\textfont0=\ninerm \scriptfont0=\sixrm \scriptscriptfont0=\fiverm
\textfont1=\ninei \scriptfont1=\sixi \scriptscriptfont1=\fivei
\textfont2=\ninesy \scriptfont2=\sixsy \scriptscriptfont2=\fivesy
\textfont\itfam=\ninei \def\it{\fam\itfam\nineit}\def\sl{\fam\slfam\ninesl}%
\textfont\bffam=\ninebf \def\bf{\fam\bffam\ninebf}\rm}
%
%
\def\noblackbox{\overfullrule=0pt}
\hyphenation{anom-aly anom-alies coun-ter-term coun-ter-terms}
\def\inv{^{\raise.15ex\hbox{${\scriptscriptstyle -}$}\kern-.05em 1}}

\def\Dsl{\,\raise.15ex\hbox{/}\mkern-13.5mu D} 
\def\dsl{\raise.15ex\hbox{/}\kern-.57em\partial}

\def\lspace{\ifx\answ\bigans{}\else\qquad\fi}
\def\lbspace{\ifx\answ\bigans{}\else\hskip-.2in\fi} 
\def\boxeqn#1{\vcenter{\vbox{\hrule\hbox{\vrule\kern3pt\vbox{\kern3pt
	\hbox{${\displaystyle #1}$}\kern3pt}\kern3pt\vrule}\hrule}}}
\def\mbox#1#2{\vcenter{\hrule \hbox{\vrule height#2in
		\kern#1in \vrule} \hrule}}  
%

\def\darr#1{\raise1.5ex\hbox{$\leftrightarrow$}\mkern-16.5mu #1}

\def\roughly#1{\raise.3ex\hbox{$#1$\kern-.75em\lower1ex\hbox{$\sim$}}}

\global\newcount\subsubsecno \global\subsubsecno=0
\def\subsubsec#1{\global\advance\subsubsecno by1%
{\toks0{#1}\message{(\the\secno\the\subsecno\the\subsubsecno. \the\toks0)}}%
\ifnum\lastpenalty>9000\else\bigbreak\fi
\noindent{\it\hyperdef\hypernoname{subsubsection}{\the\secno.\the\subsecno\the\subsubsecno}%
{\the\secno.\the\subsecno.\the\subsubsecno.} #1}
\par\nobreak\medskip\nobreak}
\def\boxit#1{\vbox{\hrule\hbox{\vrule\kern8pt
\vbox{\hbox{\kern8pt}\hbox{\vbox{#1}}\hbox{\kern8pt}}
\kern8pt\vrule}\hrule}}
\def\mathboxit#1{\vbox{\hrule\hbox{\vrule\kern8pt\vbox{\kern8pt
\hbox{$\displaystyle #1$}\kern8pt}\kern8pt\vrule}\hrule}}
\def\slashchar#1{\setbox0=\hbox{$#1$}           
   \dimen0=\wd0                                 
   \setbox1=\hbox{/} \dimen1=\wd1               
   \ifdim\dimen0>\dimen1                        
      \rlap{\hbox to \dimen0{\hfil/\hfil}}      
      #1                                        
   \else                                        
      \rlap{\hbox to \dimen1{\hfil$#1$\hfil}}   
      /                                         
   \fi}
\def\sqr#1#2{{\vcenter{\vbox{\hrule height.#2pt
         \hbox{\vrule width.#2pt height#1pt \kern#1pt
            \vrule width.#2pt}
         \hrule height.#2pt}}}}


\input amssym.def
\input amssym.tex
\noblackbox
\baselineskip=14.5pt

\def\comment#1{{}}

\def\ap{\alpha'}

\def\al{\alpha}

\newif\ifnref

\nreffalse

\input epsf

\def\figin{\epsfcheck\figin}\def\figins{\epsfcheck\figins}
\def\epsfcheck{\ifx\epsfbox\UnDeFiNeD
\message{(NO epsf.tex, FIGURES WILL BE IGNORED)}
\gdef\figin##1{\vskip2in}\gdef\figins##1{\hskip.5in}
\else\message{(FIGURES WILL BE INCLUDED)}%
\gdef\figin##1{##1}\gdef\figins##1{##1}\fi}
\def\DefWarn#1{}
\def\figinsert{\goodbreak\midinsert}  
\def\ifig#1#2#3{\DefWarn#1\xdef#1{Fig.~\the\figno}
\writedef{#1\leftbracket fig.\noexpand~\the\figno}%
\figinsert\figin{\centerline{#3}}\medskip\centerline{\vbox{\baselineskip12pt
\advance\hsize by -1truein\noindent\footnotefont\centerline{{\bf
Fig.~\the\figno}\ \sl #2}}}
\bigskip\endinsert\global\advance\figno by1}

\def\iifig#1#2#3#4{\DefWarn#1\xdef#1{Fig.~\the\figno}
\writedef{#1\leftbracket fig.\noexpand~\the\figno}%
\figinsert\figin{\centerline{#4}}\medskip\centerline{\vbox{\baselineskip12pt
\advance\hsize by -1truein\noindent\footnotefont\centerline{{\bf
Fig.~\the\figno}\ \ \sl #2}}}\smallskip\centerline{\vbox{\baselineskip12pt
\advance\hsize by -1truein\noindent\footnotefont\centerline{\ \ \ \sl #3}}}
\bigskip\endinsert\global\advance\figno by1}


\def\h {{1\over 2}}

\def\ov {\overline}
\def\o {\over}
\def\fc#1#2{{#1 \o #2}}

\def\IC{{\bf C}}\def\IR{ {\bf R}}


\def\br{\hfill\break}

\def\lf {\left}
\def\ri {\right}

\def\re {{\rm Re}}
\def\im {{\rm Im}}

 \def\Oc {{\cal O}}

\def\ceiling#1{\lceil#1\rceil}



\def\IH{{\bf H}_+}

\lref\StiebergerVYA{
  S.~Stieberger and T.R.~Taylor,
``Disk Scattering of Open and Closed Strings (I),''
Nucl.\ Phys.\ B {\bf 903}, 104 (2016).
[arXiv:1510.01774 [hep-th]].
}

\lref\StiebergerHZA{
  S.~Stieberger and T.~R.~Taylor,
  ``Superstring Amplitudes as a Mellin Transform of Supergravity,''
Nucl.\ Phys.\ B {\bf 873}, 65 (2013).
[arXiv:1303.1532 [hep-th]];\hfil\break
 S.~Stieberger and T.~R.~Taylor,
 ``Superstring/Supergravity Mellin Correspondence in Grassmannian Formulation,''
Phys.\ Lett.\ B {\bf 725}, 180 (2013).
[arXiv:1306.1844 [hep-th]].
}

\lref\StiebergerWEA{
  S.~Stieberger,
``Closed superstring amplitudes, single-valued multiple zeta values and the Deligne associator,''
J.\ Phys.\ A {\bf 47}, 155401 (2014).
[arXiv:1310.3259 [hep-th]].
}

\lref\StiebergerHBA{
  S.~Stieberger and T.R.~Taylor,
 ``Closed String Amplitudes as Single-Valued Open String Amplitudes,''
Nucl.\ Phys.\ B {\bf 881}, 269 (2014).
[arXiv:1401.1218 [hep-th]].
}

\lref\KleissNE{
  R.~Kleiss and H.~Kuijf,
  ``Multi - Gluon Cross-sections and Five Jet Production at Hadron Colliders,''
Nucl.\ Phys.\ B {\bf 312}, 616 (1989)..
}

\lref\BernQJ{
  Z.~Bern, J.J.M.~Carrasco and H.~Johansson,
``New Relations for Gauge-Theory Amplitudes,''
Phys.\ Rev.\ D {\bf 78}, 085011 (2008).
[arXiv:0805.3993 [hep-ph]].
}

\lref\StiebergerHQ{
  S.~Stieberger,
``Open \& Closed vs. Pure Open String Disk Amplitudes,''
[arXiv:0907.2211 [hep-th]].
}

\lref\stnew{
  S.~Stieberger and T.R.~Taylor,
  ``Disk Scattering of Open and Closed Strings (II),''
  in preparation.}
  
\lref\ChiodaroliRDG{
 M.~Chiodaroli, M.~G\"unaydin, H.~Johansson and R.~Roiban,
  ``Scattering amplitudes in $ {\cal N}=2 $ Maxwell-Einstein and Yang-Mills/Einstein supergravity,''
JHEP {\bf 1501}, 081 (2015).
[arXiv:1408.0764 [hep-th]].
}

\lref\KawaiXQ{
  H.~Kawai, D.C.~Lewellen and S.H.H.~Tye,
``A Relation Between Tree Amplitudes Of Closed And Open Strings,''
  Nucl.\ Phys.\  B {\bf 269}, 1 (1986).
}
\lref\MasonJY{
  L.J.~Mason and D.~Skinner,
  ``Gravity, Twistors and the MHV Formalism,''
Commun.\ Math.\ Phys.\  {\bf 294}, 827 (2010).
[arXiv:0808.3907 [hep-th]].}

\lref\StiebergerKIA{
  S.~Stieberger and T.R.~Taylor,
``Subleading Terms in the Collinear Limit of Yang-Mills Amplitudes,''
[arXiv:1508.01116 [hep-th]], to appear in Phys.\ Lett.\ B.
}

\lref\StiebergerTE{
  S.~Stieberger and T.R.~Taylor,
``Multi-Gluon Scattering in Open Superstring Theory,''
Phys.\ Rev.\ D {\bf 74}, 126007 (2006).
[hep-th/0609175].
}

\lref\notation{M.L.~Mangano and S.J.~Parke,
``Multiparton amplitudes in gauge theories,''
Phys. Rept.  {\bf 200}, 301 (1991).
[hep-th/0509223];\br
L.J.~Dixon,
  ``Calculating scattering amplitudes efficiently,''
in Boulder 1995, QCD and beyond 539-582.
[hep-ph/9601359].}

\lref\StiebergerTE{
  S.~Stieberger and T.R.~Taylor,
``Multi-Gluon Scattering in Open Superstring Theory,''
Phys.\ Rev.\ D {\bf 74}, 126007 (2006).
[hep-th/0609175].
}

\lref\BjerrumBohrRD{
  N.E.J.~Bjerrum-Bohr, P.H.~Damgaard and P.~Vanhove,
``Minimal Basis for Gauge Theory Amplitudes,''
Phys.\ Rev.\ Lett.\  {\bf 103}, 161602 (2009).
[arXiv:0907.1425 [hep-th]].
}

\lref\CachazoNSA{
  F.~Cachazo, S.~He and E.~Y.~Yuan,
  ``Einstein-Yang-Mills Scattering Amplitudes From Scattering Equations,''
JHEP {\bf 1501}, 121 (2015).
[arXiv:1409.8256 [hep-th]].
}

\lref\BernSV{
  Z.~Bern, L.J.~Dixon, M.~Perelstein and J.S.~Rozowsky,
``Multileg one loop gravity amplitudes from gauge theory,''
Nucl.\ Phys.\ B {\bf 546}, 423 (1999).
[hep-th/9811140].
}

\lref\StiebergerCEA{
  S.~Stieberger and T.R.~Taylor,
``Graviton as a Pair of Collinear Gauge Bosons,''
Phys.\ Lett.\ B {\bf 739}, 457 (2014).
[arXiv:1409.4771 [hep-th]].
}

\lref\StiebergerQJA{
  S. Stieberger and T.R.~Taylor,
``Graviton Amplitudes from Collinear Limits of Gauge Amplitudes,''
Phys.\ Lett.\ B {\bf 744}, 160 (2015).
[arXiv:1502.00655 [hep-th]].
}

\lref\KlebanovNI{
  I.R.~Klebanov and L.~Thorlacius,
``The Size of p-branes,''
Phys.\ Lett.\ B {\bf 371}, 51 (1996).
[hep-th/9510200].
}

\lref\HashimotoKF{
  A.~Hashimoto and I.~R.~Klebanov,
``Decay of excited D-branes,''
Phys.\ Lett.\ B {\bf 381}, 437 (1996).
[hep-th/9604065].
}

\lref\GubserWT{
  S.S.~Gubser, A.~Hashimoto, I.R.~Klebanov and J.M.~Maldacena,
``Gravitational lensing by $p$-branes,''
Nucl.\ Phys.\ B {\bf 472}, 231 (1996).
[hep-th/9601057].
}

\lref\DrummondVQ{
  J.M.~Drummond, J.~Henn, G.P.~Korchemsky and E.~Sokatchev,
  ``Dual superconformal symmetry of scattering amplitudes in N=4 super-Yang-Mills theory,''
Nucl.\ Phys.\ B {\bf 828}, 317 (2010).
[arXiv:0807.1095 [hep-th]].
}

\Title{\vbox{\rightline{MPP--2016--140}
}}
{\vbox{\centerline{New Relations for Einstein--Yang--Mills Amplitudes}
}}
\medskip
\centerline{Stephan Stieberger$^a$ and Tomasz R. Taylor$^b$}
\bigskip
\centerline{\it $^a$ Max--Planck--Institut f\"ur Physik}
\centerline{\it Werner--Heisenberg--Institut, 80805 M\"unchen, Germany}
\medskip
\centerline{\it  $^b$ Department of Physics}
\centerline{\it  Northeastern University, Boston, MA 02115, USA}

\vskip15pt

\medskip
\bigskip\bigskip\bigskip
\centerline{\bf Abstract}
\vskip .2in
\noindent

\noindent
We obtain new relations between Einstein--Yang--Mills (EYM) amplitudes involving $N$ gauge bosons plus a single graviton and pure Yang--Mills amplitudes involving $N$ gauge bosons plus one additional vector boson inserted in a way typical for a gauge boson of a ``spectator'' group commuting with the group associated to  original $N$ gauge bosons. We show that such EYM amplitudes satisfy U(1) decoupling relations similar to Kleiss--Kuijf relations for Yang--Mills amplitudes. We consider a D--brane embedding of EYM amplitudes in the framework of disk amplitudes involving open and closed strings. 
A new set of monodromy relations is derived for mixed open--closed amplitudes  with one closed string inserted on the disk world--sheet and a number of open strings at the boundary.
These relations allow expressing the latter in terms of pure open string amplitudes
and, in the field--theory limit, they yield the U(1) decoupling relations for EYM amplitudes.

\Date{}
\noindent
\goodbreak
\break
\newsec{Introduction}

\noindent Tree--level Einstein--Yang--Mills (EYM) amplitudes in Yang--Mills (YM) theory with the energy-momentum tensor (minimally) coupled to Einstein's gravity offer a useful laboratory for studying the nature of gravitons. In this paper, we discuss the case of a single graviton emitted (or absorbed) in a scattering process involving a number of vector bosons. We also discuss a superstring framework in which gauge bosons appear as massless states of open strings on D--branes and the graviton originates from a closed string. In this case, the disk amplitudes describing such scattering processes include non--minimal couplings generated by the tower of open string states populating Regge trajectories with the slope $\alpha'$. At low energies, with all invariant mass scales $M^2\alpha'\to 0$ one recovers EYM theory with the gravitational coupling $\kappa\equiv 8\pi G_N\sim \alpha'$.

The paper is organized as follows. In Section 2, we start from the recent observation \refs{\StiebergerKIA,\StiebergerVYA} that one--graviton EYM amplitudes can be expressed as  linear combinations of pure YM amplitudes with two additional vector bosons inserted instead of the graviton. We take the soft limit in which one of these particles gets no momentum while the second one takes over the whole momentum from the graviton. In this way, we obtain new relations between EYM  amplitudes involving $N$ gauge bosons plus a single graviton and pure YM amplitudes involving $N$ gauge bosons plus one additional vector boson. We discuss relations between distinct partial amplitudes, in particular the implications of U(1) decoupling similar to the case of pure gauge amplitudes. In Section 3, we move to full--fledged superstring disk amplitudes with an arbitrary  number of open strings attached at the boundary and a single closed string inserted on the world--sheet. By using analytic continuation, we obtain contour integral representations of the amplitudes and discuss their monodromy properties. As a result, we obtain a new set of monodromy relations for mixed open--closed amplitudes. The new element is the existence of a novel string ``tube'' contribution. In section 4, we discuss the $\alpha'\to 0 $ limit of these monodromy relations and show that U(1) decoupling does indeed follow in the EYM field-theoretical limit.

\newsec{New Relations between Einstein-Yang-Mills and Yang-Mills Amplitudes}

 In Refs. \refs{\StiebergerKIA,\StiebergerVYA}, we showed that a class of tree--level EYM amplitudes describing decays of a single graviton into $N{-}2$ gauge bosons can be written as linear combinations of pure gauge, $N$--particle amplitudes in which the graviton is replaced by the pair $\{N{-}1,N\}$ of gauge bosons. The momenta,
\eqn\conf{p_{N-1}=x\ P~, \qquad p_{N}=(1-x)\ P\ ,}
are collinear, with one gauge boson carrying the fraction $x$ of the graviton momentum $P$, and the second the remaining $1-x$. These gauge bosons carry identical helicities which
add up to graviton's $+2$ or $-2$. The relation reads
\eqnn\FINALL{
$$\eqalignno{\quad A_{\rm EYM}(1,2,\dots&,N{-}2;P^{\pm\! \pm})
=&\FINALL\cr
\quad={\kappa\,x(1-x)\over g^2}&\Biggl\{ \!\sum_{l=2}^{\ceiling{\fc{N}{2}}-1}\!\sum_{i=2}^l
\Big(\sum_{j=i}^ls_{jP}\Big)\, A_{\rm YM}(1,\ldots,i{-}1,N^{\pm},i,\dots,l,N{-}1^\pm,l{+}1,\ldots,N{-}2)\cr
\,\,+\sum_{l=\ceiling{\fc{N}{2}}}^{N-3}&\sum_{i=l+1}^{N-2}
\Big(\!\sum_{j=l+1}^is_{jP}\Big)\ A_{\rm YM}(1,\ldots,l,N{-}1^\pm,l{+}1,\ldots,i,N^\pm,i{+}1,\ldots,N{-}2)\Biggr\} ,}
$$}
$\!\!$where $\kappa$ and $g$ are the gravitational and gauge coupling constants, respectively\foot{$\ceiling{{N\over 2}}$ is  the smallest integer greater than or equal to ${N\over 2}$. Since the graviton is identified by its momentum $P$, we can skip in the following the EYM and YM labelings of the amplitudes.}.  On the left hand side, we have a mixed gauge--gravitational amplitude involving a single graviton of momentum $P$, helicity $+2$ or $-2$, as indicated by the superscript, and  $N{-}2$ gluons. This amplitude is associated to a single trace color factor with the respective gluon ordering.
On the right hand side, we have a linear combination of pure gauge, partial amplitudes weighted by the kinematic invariants $s_{jP}=2p_jP$. Here, the graviton is replaced by two gluons in the collinear configuration $p_{N-1}=xP,~  p_{N}=(1-x)P$.
 Note that on the right hand side, $N{-}1$ and $N$ are never adjacent, therefore the respective YM   amplitudes do not contain collinear singularities.
In particular, for the lowest multiplicities $N=5,6,7$,
Eq. \FINALL\ yields:
\eqnn\three
\eqnn\four
\eqnn\five
$$\eqalignno{A(1,2,3;P^{\pm\! \pm})&={\kappa\,x(1-x)\over g^2}\,s_{2P} A(1,5^\pm,2,4^\pm,3)\ ,&\three\cr
A(1,2,3,4;P^{\pm\! \pm})&={\kappa\,x(1-x)\over g^2}\Big\{s_{2P} A(1,6^\pm,2,5^\pm,3,4)+s_{4P}\ A(1,2,3,5^\pm,4,6^\pm)\Big\}\ ,&\four\cr
A(1,2,3,4,5;P^{\pm\! \pm})&={\kappa\,x(1-x)\over g^2}\Big\{s_{2P} A(1,7^\pm,2,6^\pm,3,4,5)+s_{3P} A(1,2,7^\pm,3,6^\pm,4,5)\cr
&+(s_{3P}+s_{2P}) A(1,7^\pm,2,3,6^\pm,4,5)+s_{5P} A(1,2,3,4,6^\pm,5,7^\pm)\Big\}\ .\ \ \ \ \ \ \ &\five}$$

Since the graviton amplitude of Eq. \FINALL\ does not refer to any particular $x$, the r.h.s.\ must be
$x$--independent. This observation allows taking the limit $x\to 0$, {\it i.e}.\ $p_{N-1}=xP \to 0$ and
$p_{N}=(1-x)P \to P$. It is the ``soft'' limit for the $(N-1)$--th gluon, in which the $N$--th one carries  the whole graviton momentum. In our case, only the leading soft singularity \notation\ contributes in $x\to 0$ limit:
\eqnn\limm
\eqnn\limn
$$\eqalignno{\lim_{x\to 0}x(1-x)\ A(\dots,m,xP^+\! ,n,\dots)& ~=~ g {\langle mn \rangle\over\langle mP\rangle\langle Pn\rangle}\, A(\dots,m,n,\dots)\ , &\limm\cr
\lim_{x\to 0}x(1-x)\ A(\dots,m,xP^-\! ,n,\dots)& ~=~\, g {[ mn ]\over[ mP][ Pn]}~ A(\dots,m,n,\dots)\ . &\limn}$$
As a result, after some particle relabelling, we obtain
\eqnn\singlep
\eqnn\singlem
$$\eqalignno{A(1,2,\dots N;P^{+\! +}) & ~=~{\kappa\over g}\ \sum_{l=2}^{N-1}{\langle 1|x_l|P]\over\langle 1P\rangle}~
A(1,2,\dots,l,P^+\! ,l{+}1,\dots, N)\ , &\singlep\cr
A(1,2,\dots N;P^{-\! -}) & ~=~{\kappa\over g}\ \sum_{l=2}^{N-1}{[ 1|x_l|P\rangle\over [ 1P]}~
A(1,2,\dots,l,P^-\! ,l{+}1,\dots, N)\ , &\singlem}$$
where we used the standard helicity notation  \notation, with the dual conformal coordinate
\DrummondVQ\ defined as $x_l=\sum_{k=1}^{l}p_k$. Of course, we could also consider the limit $x\to 1$ in which the other gluon becomes soft. This yields expressions that can be shown by using BCJ \BernQJ\ relations to be equivalent to the r.h.s.\ of Eqs. \singlep\ and \singlem.

There is another, particularly interesting way of rewriting our result
\eqn\weps{A(1,2,\dots N;P^{\pm\! \pm}) ~=~{\kappa\over g}\sum_{l=1}^{N-1}(\epsilon^{\pm}_P\!\cdot\! x_l)~
A(1,2,\dots,l,P^{\pm}\! ,l{+}1,\dots, N)\ ,}
where $\epsilon^{\pm}_P$ are the spin 1 polarization vectors of a gauge boson with momentum $P$. Here, the gauge invariance of the r.h.s.\ follows from BCJ relations. With the choice of $p_1$ as the reference vector, Eq. \weps\  yields Eqs. \singlep\ and \singlem. Written in this way, Eq. \weps\ holds in any number of dimensions. On the r.h.s.,  the graviton is inserted into partial gauge amplitudes in the same way as a vector boson of a ``spectator'' group commuting with the group associated to $N$ gauge bosons. The factors ($\epsilon_P\cdot x_l$), which are typical of a gauge boson coupled to a scalar line, are not unfamiliar to gravitational amplitudes: they have already appeared in the Mason--Skinner \MasonJY\ representation of  multi--graviton MHV amplitudes. For that reason, we expect that some equations similar to \weps\ hold also for EYM amplitudes involving more than one graviton.

EYM amplitudes have been studied before in the framework of scattering equations \CachazoNSA. It would be very interesting to see how a rather complicated formula written in \CachazoNSA\ can reproduce Eq. \weps.

By using the explicit representation of Eqs. \singlep, \singlem\ and \weps\ and the properties of pure gauge amplitudes it is easy to verify that one--gluon EYM amplitudes have the following reflection property:
\eqn\pari{A(1,2,\dots N;P)=(-1)^N A(N,\dots,2, 1;P)\ .}
Furthermore, they satisfy exactly the same U(1) decoupling relations as the well--known Kleiss-Kuijf relations \KleissNE\ that hold in the absence of the graviton:
\eqnn\kkmix
$$\eqalignno{A(1,2,3,4,\dots, N;P)+A(1,3,2,4,\dots, N;P) &+A(1,3,4,2,\dots, N;P)\,+&\kkmix\cr
\dots &+ A(1,3,4,\dots, N,2;P) ~=~ 0\ .&}$$
The above relations reflect the fact that the respective
Feynman diagrams can be constructed by inserting the graviton on all possible internal and external lines of $N$--gluon diagrams.

In the following section we will discuss EYM amplitudes in the framework of superstring theory. We will derive monodromy relations for disk amplitudes with a single closed string inserted on the world--sheet. In this context,
Eq. \kkmix\ will appear in the field theory limit of highly non--trivial monodromy relations on the string world--sheet.

\newsec{World--sheet  monodromy relations for mixed string amplitudes}

In \StiebergerHQ\ tree--level string amplitudes involving both open and closed strings have  been expressed as linear combinations of pure open string amplitudes.
This correspondence gives a  relation between amplitudes involving both gluons and gravitons and pure gauge amplitudes at tree--level \StiebergerCEA\ with interesting consequences
for constructing  gravity amplitudes from gauge amplitudes \StiebergerQJA.
In this section by applying world--sheet string techniques
we derive new algebraic identities between  mixed string amplitudes involving both open and closed strings and pure open string amplitudes.
In the field--theory limit these relations give rise to amplitude relations between EYM amplitudes. We shall work at the leading order (tree--level) in string perturbation theory.

Tree--level amplitudes involving both open and closed strings are described by a disk world--sheet,
which is an oriented manifold with one boundary.
The latter can be mapped to the upper half plane:
\eqn\upper{
\IH=\{z\in \IC\ |\ \im(z)\geq 0\ \}\ .}
Open string vertex operator $V_o(x)$ insertions $x$ are placed at the boundary of the disk and closed string
positions $z$ in the bulk.
The techniques for evaluating generic disk integrals involving both open and  closed string states have been developed in \refs{\StiebergerHQ,\StiebergerCEA,\StiebergerVYA}.
The amplitudes can be decomposed as certain linear combinations of pure open string amplitudes.
Formally, the computation of disk amplitudes involving both open and closed strings is reduced
to considering the monodromies on the complex sphere.

Scattering amplitudes of open and closed strings describe the couplings of brane and bulk fields thus probing the effective  D--brane action.
In the following we shall consider disk amplitudes with one bulk and $N-2$ boundary operators\foot{Disk amplitudes with an arbitrary number of  bulk and  boundary operators will be considered in~\stnew.}.
This yields the leading order amplitude for either the absorption of a closed string by a D--brane
or the decay of an excited D--brane into a massless closed string state and the unexcited D--brane
\refs{\KlebanovNI,\GubserWT,\HashimotoKF}.
Open string vertices  with momenta $p_i,~i=1,\dots,N{-}2$ are inserted on the real axis of \upper\ at
$x_i\in\IR$,  while a single closed string vertex operator  is inserted at complex $z\in\IH$.
For the latter we assume different left-- and right--moving space--time momenta $q_1$ and $q_2$, satisfying the massless on--shell condition $q_i^2=0$, respectively.  This is the most general setup for scattering both open and closed strings in the presence of D--branes and orientifold planes. We refer the reader to \StiebergerVYA\ for further details.

We shall discuss the mixed amplitude
\eqn\Start{
A(1,2,\ldots,N-2;q_1,q_2)}
involving $N-2$ open and one closed string state with different
left-- and right--moving space--time momenta $q_1,q_2$. The amplitude \Start\ has been computed in \StiebergerVYA.
If the closed string momenta are left--right symmetric, i.e. $q_1=q_2$  reflection symmetry is furnished in the amplitude \Start\ as\foot{Here, complex conjugation $\ast$ acts at the world--sheet integral, specified in Eq. (3.4), while kinematical factors are unaffected.}:
\eqn\reflection{
A(1,2,\ldots,N-2;q,q)=(-1)^N\ A(1,N-2,\ldots,2;q,q)^\ast\ .}
However, for generic momenta $q_1$ and $q_2$ the relation symmetry \reflection\
does not hold. To be most general this is what we shall assume in the following.
In \Start\ any kinematical factor is multiplied by some form factor described by a complex integral of the form \StiebergerVYA
\eqn\GENERIC{\eqalign{
&F(1,2,\ldots,N-2;q_1,q_2)\cr
&\hskip0.25cm= V_{\rm CKG}^{-1}\ \delta\Big(\sum_{i=1}^{N-2}p_i+q_1+q_2\Big)\hskip-0.25cm\int\limits_{x_1<\dots<x_{N{-}2}}\hskip-0.2cm\lf(\prod_{i=1}^{N-2} dx_i\ri)
\prod_{1\leq r<s\leq N-2}|x_r-x_s|^{2\alpha'p_rp_s}\ (x_r-x_s)^{n_{rs}}\cr
&\hskip0.25cm\times \int_{\IH} d^2z\ (z-\ov z)^{2\al'q_1q_2+n} \ \prod_{i=1}^{N-2}\ (x_i-z)^{2\al'p_iq_1+n_i}\ (x_i-\ov z)^{2\al'p_iq_2+\ov n_i}\ ,}}
where we included the momentum--conserving (along the D-brane world--volume)
\eqn\totalclosed{
\sum_{i=1}^{N-2}p_i+q_1+q_2=0}
delta function and divided by the volume $V_{\rm CKG}$ of the conformal Killing group.
To be specific, we focus on the amplitude associated to
one particular Chan-Paton factor (partial amplitude), ${\rm Tr}(T^1T^2\dots T^{N-2})$, with the real iterated integral over the domain $x_1<x_2<\dots<x_{N{-}2}$. Note, that in \GENERIC, the momenta
$q_1$ and $q_2$ are assumed to be unrelated.
Finally, the powers $n_{rs},~n_i,~\ov n_i,~ n$ are some integer numbers specified by the kinematics under consideration.

In the following we shall discuss monodromy properties arising from that part of \GENERIC\
which has non--integer exponents.  The branching is caused by the factors
$(x_r-x_s)^{2\ap p_rp_s}, (x_i-z)^{2\al'p_iq_1}, (x_i-\ov z)^{2\al'p_iq_2}$ and $(z-\ov z)^{2\al'q_1q_2}$. On the other hand, the monodromy properties are not affected by the choice of the integer exponents.
Hence, the choice of the latter will not enter in the next steps
and the following results are completely independent on the latter. As a consequence our monodromy properties, which can be stated for any given kinematics referring to particular choice of integers, hold
for the full amplitude \Start.

Without any restriction in \GENERIC\ we may assume $x_1=-\infty$ and then consider the real integration w.r.t. e.g. $x_2$.
Analytically continuing the $x_2$--integration to the whole complex plane and choosing the integration w.r.t. $x_2$ along the  contour integral depicted in Fig. 1 gives rise to  the following relation
\eqnn\DUAL{
$$\eqalignno{
A(1,2,&\ldots,N-2;q_1,q_2)+e^{-i\pi s_{23}}\ A(1,3,2,\ldots,N-2;q_1,q_2)\cr
&+e^{-i\pi( s_{23}+ s_{24})}\ A(1,3,4,2,\ldots,N-2;q_1,q_2)&\DUAL\cr
&+\ldots+e^{-i\pi( s_{23}+ s_{23}+\ldots+ s_{2,N-2})}\ A(1,3,\ldots,N-2,2;q_1,q_2)=T(3,\ldots,N-2)\ ,}$$}
with
\eqn\Mandel{
s_{ij}\equiv s_{i,j}=2\ap\ k_ik_j\ ,}
and the $N$ open string momenta $k_r=p_r,\ r=1,\ldots,N-2$,\ $k_{N-1}=q_1$ and $k_N=q_2$  \StiebergerVYA.
\ifig\plahte{Contour integral in the complex $x_2$--plane.}{\epsfxsize=0.8\hsize\epsfbox{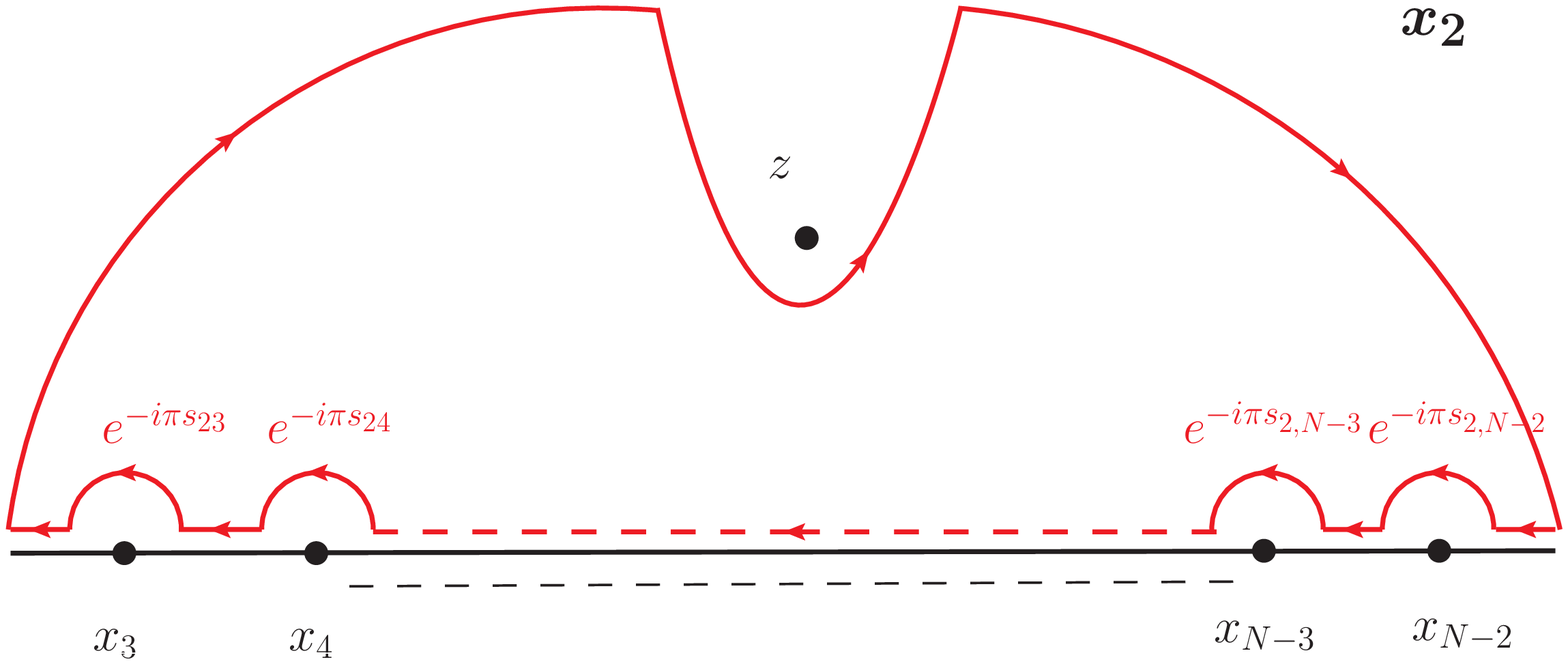}}
\noindent
Eq. \DUAL\ gives rise to a new set of monodromy relations for mixed open--closed amplitudes.
The new element is the existence of a novel string ``tube'' contribution $T(3,\ldots,N-2)$  to be specified below.
Without this contribution  the relation \DUAL\ boils down to the open string monodromy relations discussed in \refs{\StiebergerHQ,\BjerrumBohrRD}.
The choice of the contour along the real axis accommodates the correct branch of the integrand
and gives rise to a phase factor each
time when encircling one open string vertex position $x_j,\ j=3,\ldots,N-2$.
Note that the phases, which are independent on the integers
$n_{rs},n_i,\ov n_i, n$ do not depend on the particular values of  integration variables, but only on the ordering of $x_2$ with respect to the remaining original  vertex positions.
On the other hand, the semicircle can be deformed to infinity by taking into account the infinite tube around the closed string position $z$. At infinity the integrand
behaves as $x_2^{-2h_2}$ with $h_2$ the conformal weight of the vertex
operator $V_o(x_2)$. Since we consider physical states, we have $h_2=1$
and thus there is no contribution from the semicircle. On the other hand, there is  one contribution  from the infinite long tube encircling the closed string position $z$. In fact, after fixing the latter to $z=i$ this amount can be written as\foot{Note, that we have dismissed the integer exponents $n_{rs},~n_i,~\ov n_i,~ n$, which may easily be reinstalled, cf. the discussion after Fig. 1.}:
\eqnn\Tube{
$$\eqalignno{
T(3,&\ldots,N-2)=\delta\Big(\sum_{i=1}^{N-2}p_i+q_1+q_2\Big)\ \sin(\pi s_{2,N-1})\ e^{-i\pi s_{1,N}}\cr
&\times\int_1^\infty dy\ |y-1|^{s_{2,N-1}}\ |y+1|^{s_{2,N}}
\hskip-0.5cm\int\limits_{x_3<\ldots<x_{N-2}}\lf(\prod_{i=3}^{N-2} dx_i\ri)
\prod_{3\leq r<s\leq N-2}|x_r-x_s|^{s_{rs}}\cr
&\times  \prod_{j= 3}^{N-2} |x_j-iy|^{s_{2,j}}\  |x_j-i|^{s_{j,N-1}}\
|x_j+i|^{s_{j,N}}\ .&\Tube}$$}

To familiarize with \Tube\ let us first discuss the case $N=5$ and compute $T(3)$ contributing at the r.h.s. of \DUAL. Eventually, for this case the latter yields the following relation:
\eqn\DUALf{
A(1,2,3;q_1,q_2)+e^{-i\pi s_{23}}\ A(1,3,2;q_1,q_2)=-2i\ e^{-i\pi s_{51}}\ \sin(\pi s_{24})\ \sin(\pi s_{35})\ A(1,2,4,5,3)\ .}
The r.h.s. of \DUALf\ stems from the infinite tube \Tube:
\eqn\Tubef{\eqalign{
T(3)=\sin(\pi s_{24})\ e^{-i\pi s_{51}}& \int_1^\infty dy\ |y-1|^{s_{24}}\ |y+1|^{s_{25}}\cr
&\times \int_{-\infty}^\infty dx_3\ |x_3-iy|^{s_{23}}\ |x_3-i|^{s_{34}}\ |x_3+i|^{s_{35}}\ .}}
The only remaining open string position to be integrated along the real axis is $x_3$.
This integration can conveniently be  deformed  to the imaginary axis
along the contour $C$ depicted in Fig.~2.
\ifig\tubef{Contour integral in the complex $x_3$--plane.}{\epsfxsize=0.5\hsize\epsfbox{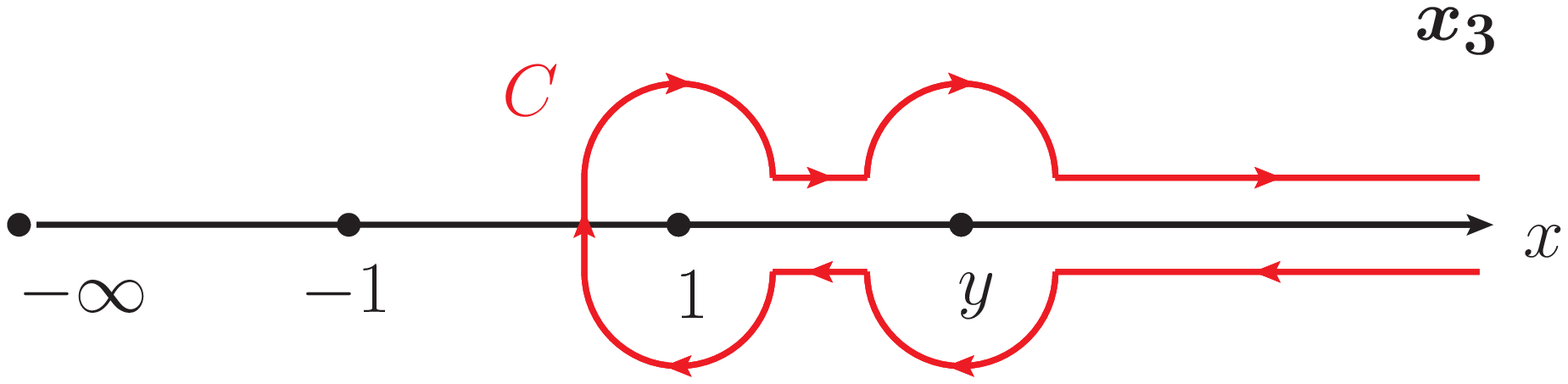}}
\noindent
and the expression \Tubef\ becomes ($x_3=ix$):
\eqn\Tubeff{
T(3)=\sin(\pi s_{24})\ e^{-i\pi s_{51}}\ \int_1^\infty dy\ |y-1|^{s_{24}}\ |y+1|^{s_{25}}
\int_C dx\ |x-y|^{s_{23}}\ |x-1|^{s_{34}}\ |x+1|^{s_{35}}\ .}
This integral resembles a generic  open string integral involving five open strings
\eqn\Tubefff{
T(3)=2i\ e^{-i\pi s_{51}}\ \sin(\pi s_{24})\
 \lf\{\ \sin(\pi s_{34})\ A(1,5,4,3,2)+\sin[\pi(s_{23}+s_{34})]\
A(1,5,4,2,3)\ \ri\}\ ,}
which agrees with the r.h.s. of \DUALf\ thanks to pure open string monodromy relations \refs{\StiebergerHQ,\BjerrumBohrRD}.

Let us now move on to the  case $N= 6$.
As in the case of $N=5$ in \Tube\ we would like to deform the remaining two real integrations w.r.t. $x_3$ and $x_4$ to the imaginary axis.
However, due to the iterated integral structure respecting
$x_3<x_4$ this procedure is not as straightforward as in the $N=5$ case. On the other hand, the tube contribution \Tube\ itself satisfies monodromy relations.
By relating the iterated integrations over the open string positions $x_3,x_4$
to contours in the complex plane we find
\eqn\find{\eqalign{
T(3,4)&+e^{-i\pi s_{34}}\ T(4,3)=:R(3,4)=\sin(\pi s_{25})\ e^{-i\pi  s_{16}}
\int_1^\infty dy\ |y-1|^{s_{25}}\ |y+1|^{s_{26}}\cr
&\times\int_{C_x} dx\int_{C_z} dz\ |z-x|^{s_{34}}\ |x-y|^{s_{23}}\ |x-1|^{s_{35}}\
|x+1|^{s_{36}}\  |z-y|^{s_{24}}\ |z-1|^{s_{45}}\ |z+1|^{s_{46}}, }}
with the two contours $C_x$ and $C_z$ depicted in  Fig. 3 and the Heaviside step
function $\Theta$.
\ifig\tubef{Contour integral in the complex $x_3$-- and $x_4$--planes.}{\epsfxsize=0.5\hsize\epsfbox{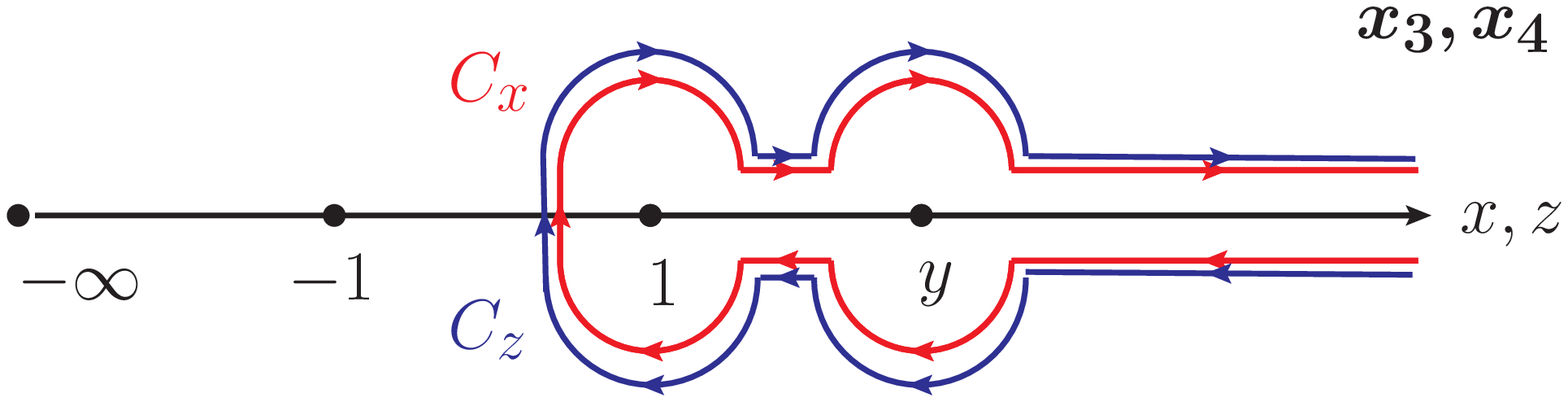}}
\noindent
Similarly to \Tubeff\ the contour integrals can be decomposed into a sum over open string  six--point amplitudes yielding the following expression:
\eqnn\Res{
$$\eqalignno{
R(3,4)&=-4\ \sin(\pi s_{25})\ e^{-i\pi  s_{16}}&\Res\cr
&\times\sum_{\sigma\in S_{3}}\prod_{j=3}^4
\sin\lf\{\pi\lf(s_{j,5}+\sum_{l=2}^{j-1}s_{jl}\ \Theta[\sigma^{-1}(j)-\sigma^{-1}(l)]\ri)\ri\}\ A(1,6,5,\sigma(2,3,4))\ .}$$}
Eventually, considering also the monodromy relation \find\ with  $3$ and $4$ permuted allows to gain an explicit expression for the tube:
\eqn\Explicit{
T(3,4)=-\fc{i}{2}\ \fc{1}{\sin(\pi s_{34})}\ \lf[\ e^{i\pi s_{34}}\ R(3,4)-R(4,3)\ \ri]\ .}
Successively considering contour deformations in the positions $x_3,\ldots,x_{N-2}$ and applying monodromy relations for \Tube\ yields
\eqn\TUBE{
\sum_{\rho \in S_{N-4}}\prod_{3\leq j<l\leq N-2}
\exp\lf\{-i\pi \Theta[\rho^{-1}(j)-\rho^{-1}(l)]\ s_{jl}\ri\}\ T(\rho(3,\ldots,N-2))=R(3,\ldots,N-2)\ }
for generic $N$ with
\eqnn\RES{
$$\eqalignno{
 R(3,\ldots,N-2)&=(2i)^{N-4}\ e^{-i\pi  s_{1,N}}\ \sin(\pi s_{2,N-1})\cr
&\times \sum_{\sigma\in S_{N-3}}\prod_{j=3}^{N-2}
\sin\lf\{\pi\lf(s_{j,N-1}+\sum_{l=2}^{j-1}s_{jl} \Theta[\sigma^{-1}(j)-\sigma^{-1}(l)]\ri)\ri\}\cr
&\times A(1,N,N-1,\sigma(2,\ldots,N-2))\ ,&\RES}$$}
which reduces to \Tubefff\ for $N=5$ and to \Res\ for $N=6$, respectively.
Considering all  equations obtained from \TUBE\ by permuting the labels $3,\ldots,N-2$ provides
$(N-4)!$ linear equations for as many unknown tube contributions $T(\rho(3,\ldots,N-2))$.
Hence,  the resulting system allows to determine the latter in terms of
\RES\ and permutations thereof. This has been demonstrated for the case $N=6$  in \Explicit.
Furthermore, for $N=7$ we find:
\eqnn\EXPLICIT{
$$\eqalignno{
T(3,4,5)&=\fc{1}{4}\ \fc{1}{\sin(\pi s_{34})\sin(\pi s_{45})\sin[\pi(s_{34}+s_{35}+s_{45})]}&\EXPLICIT\cr
&\times\Big[\ e^{i\pi (s_{34}+s_{35}+s_{45})}\ \lf\{-\sin[\pi(s_{34}+s_{45})]\ R(3,4,5)+\sin(\pi s_{34})\ R(3,5,4)\ri.\cr
&+\lf.\sin(\pi s_{45})\ R(4,3,5)\ri\}+\sin(\pi s_{34})\ R(4,5,3)\cr
&+\sin(\pi s_{45})\ R(5,3,4)-\sin[\pi(s_{34}+s_{45})]\ R(5,4,3)\ \Big]\ .}$$}

One can verify, that the relations\foot{An alternative relation other than \DUAL\ can be obtained by inserting the latter into \TUBE\ giving rise to the relation
\eqn\FINAL{
\sum_{\rho \in S_{N-3}}\prod_{2\leq j<l\leq N-2}
\exp\lf\{-i\pi \Theta[\rho^{-1}(j)-\rho^{-1}(l)]\ s_{jl}\ri\}\ A(1,\rho(2,\ldots,N-2);q_1,q_2)=R(3,\ldots,N-2)\ ,}
which gives \DUALf\ for $N=5$. Note, that the r.h.s. of \FINAL\
starts at the order $\ap^{N-3}$.} \DUAL\ and \FINAL\ are  fulfilled by the mixed amplitudes computed in \StiebergerVYA.
Of course, by analytically continuing any other open string position $x_j,\ j=3,\ldots,N-2$
we can generate further  relations of the type \DUAL, which allow for expressing all mixed subamplitudes
\Start\ in terms of pure open string amplitudes.
Note, that the proof of the relation \DUAL\ and its permutations
does neither rely on any kinematical properties of the subamplitudes, on the amount of supersymmetry nor on the space--time dimension. Moreover,  \DUAL\ holds for any type
of massless string states both from the NS and R sector.
Hence, these relations are valid in any space--time dimension $D$, for
any amount of supersymmetry and any gauge group.

\newsec{Relations for EYM subamplitudes from string world--sheet monodromies}

The field--theory limit of \DUAL\ is given by  taking in \Start\ the lowest order in $\ap$ (in the following denoted by  $A_{FT}$), and replacing the exponentials  by $e^{i\pi s_{ij}}\sim 1+\Oc(\ap)$
\eqnn\BOIL{
$$\eqalignno{
A_{FT}(1,2,&\ldots,N-2;q_1,q_2) + A_{FT}(1,3,2,\ldots,N-2;q_1,q_2)&\BOIL\cr
&+  A_{FT}(1,3,4,2,\ldots,N-2;q_1,q_2)+\ldots+  A_{FT}(1,3,\ldots,N-2,2;q_1,q_2)=0\ ,}
$$}
since the infinite tube contribution \Tube\ is a string effect of higher order. More precisely, its real part is of order
$\ap^3$, while its imaginary part is of order $\ap^2$.
For $q_1=q_2=\h P$ the real part of \Start\ provides  the corresponding amplitudes in the double cover, whose field--theory limit gives the EYM amplitudes \refs{\StiebergerCEA,\StiebergerVYA}:
\eqn\EYM{
\re\lf.A_{FT}(1,\ldots,N-2;q_1,q_2)\ri|_{ q_1=q_2=\h P}=-\h\ A_{EYM}(1,\ldots,N-2;P)\ .}
Note, that thanks to  \reflection\ the amplitude \EYM\ enjoys \pari\ . Then, \BOIL\ yields the relation
\eqnn\BOILL{
$$\eqalignno{
A_{EYM}(1,2,&\ldots,N-2;P) +A_{EYM}(1,3,2,\ldots,N-2;P)&\BOILL\cr
&+  A_{EYM}(1,3,4,2,\ldots,N-2;P)+\ldots+  A_{EYM}(1,3,\ldots,N-2,2;P)=0\ ,}
$$}
which exhibits Kleiss--Kuijf \KleissNE\ relations in the gluon sector, e.g. for $N=5$ we have
\eqn\FTf{
A_{EYM}(1,2,3;P)+A_{EYM}(1,3,2;P)=0\ ,}
while for $N=6$ we obtain:
\eqn\FTs{
A_{EYM}(1,2,3,4;P)+ A_{EYM}(1,3,2,4;P)+ A_{EYM}(1,3,4,2;P)=0\ .}
In addition to the relations \BOILL\ for the real part \EYM\ Eq. \BOIL\ also comprises the same equations for the lowest order imaginary part of \Start. 

Respecting higher orders of \DUAL\ gives rise to various equations relating field--theory amplitudes to their string corrections. In particular, that linear combination of EYM amplitudes which would vanish in YM theory (i.e.\ in the absence of graviton) as a result of BCJ relations \BernQJ\ is non--vanishing because it obtains corrections from the tube contribution \Tube.

On the other hand, by considering different permutations of \DUAL\ we can solve for
explicit expressions for EYM amplitudes in terms of the lowest order of the tube contribution \Tube, e.g. for $N=5$ this yields:
\eqn\yields{\eqalign{
s_{23}\ A_{EYM}(1,2,3,P)&=-\fc{1}{\pi}\ \im \lf.\lf\{T(3)-e^{-i\pi s_{23}}\ T(2)\ri\}\ri|_{\ap^2}\cr
&=2\pi\ \lf\{s_{24}s_{35}\ A(1,2,4,5,3)-s_{34}s_{25}\ A(1,3,4,5,2)\ri\}\cr
&=\pi\ s_{23}s_{24}\ A(1,5,2,4,3)\ .}}

\newsec{Conclusions}

According to Eqs. \singlep, \singlem\ and \weps, the single--trace amplitude involving one graviton and $N$  gauge bosons can be expressed as a linear combination of partial gauge amplitudes in which one additional gauge boson is inserted instead of the graviton. This additional particle couples in a way typical for a gauge boson of a ``spectator'' group commuting with the group gauged by the original vector bosons.

In Refs. \StiebergerHZA, we showed that at the tree--level, multi--graviton supergravity amplitudes can be mapped, by using a particular type of Mellin transformation, into full--fledged open string amplitudes in which each graviton is replaced by a vector boson. For one--graviton EYM amplitudes written in Eq. \weps, a similar transformation yields open string amplitudes with the graviton replaced by a ``spectator'' vector boson and the original vector particles substituted by scalar particles, cf.\ $(\epsilon_P\cdot x_l)$ coefficients.  Thus Mellin correspondence reduces spin by one unit at the cost of introducing Regge excitations of lower spin particles.

The results obtained in this paper, together with earlier Refs. \refs{\StiebergerCEA,\StiebergerQJA,
\ChiodaroliRDG}, color--kinematic duality \BernQJ\ and results in string theory \refs{\KawaiXQ,\StiebergerWEA,\StiebergerHBA}, combine to an intriguing collection of observations indicating the existence of some yet unknown, deep connections between
gravity and gauge theories. We believe that even more insight will be obtained by
studying more complex  multi-graviton EYM amplitudes \stnew.

\vskip1cm
\goodbreak
\leftline{\noindent{\bf Acknowledgments}}

\noindent
We are grateful to CERN Theoretical Physics Department, where parts of this work were performed, for its hospitality and financial support. St.St.\ thanks Nordita at Stockholm and TRT is grateful to LPTHE at l'Universit\'e Pierre et Marie Curie in Paris.
This material is based in part upon work supported by the National Science Foundation under Grant No.\ PHY-1314774.   Any
opinions, findings, and conclusions or recommendations expressed in
this material are those of the authors and do not necessarily reflect
the views of the National Science Foundation.

\listrefs

\end
{\it Since $p_1=q_1$ (with $\mu_1=\nu_1$) for adjacent gluons, in order to obtain a well-defined amplitude, we start from $p_1\neq q_1$ and take the limit after symmetrizing in $\{p_1,q_1\}$, thus removing the leading collinear singularity. By using BCJ relations \BernQJ, such a symmetric combination can always be rewritten in terms of manifestly finite amplitudes in which the collinear gluons appear at non--adjacent positions, {\it cf}.\ in the example discussed at the end of the paper.} {\bf Is this paragraph still an issue - only for $N=3$, so perhaps we can move this comment to the example ?}
\eqn\adef{\eqalign{A[&p,N{-}1,N{-}2,\dots,2, 1,1,2,\dots,N{-}2,N{-}1,q]\cr &\equiv
A[p,\mu={+}1;p_{N{-}1},\mu_{N{-}1};\dots;p_1,\mu_1; q_1,\nu_1;\dots;q_{N{-}1},\nu_{N{-}1};
q,\nu={-}1]~.
}}